\documentclass[prd,aps,preprint,showpacs,floatfix]{revtex4}
\usepackage{graphicx}
\usepackage{epsf}
\usepackage{pslatex}

\def\mpi2{m_\pi^2}
\def\mK2{m_K^2}

\newcommand{\bea}{\begin{eqnarray}}
\newcommand{\eea}{\end{eqnarray}}
\newcommand{\be}{\begin{equation}}
\newcommand{\ee}{\end{equation}}
\newcommand{\nn}{\nonumber}
\newcommand{\VEV}[1]{\left\langle #1\right\rangle}

\newcommand{\A}{{\cal A}}
\newcommand{\B}{{\cal B}}

\newcommand{\h}{ h}
\newcommand{\el}{ l}

\newcommand{\C}{{\cal C}}

\newsavebox{\DERIVBOXZLM}
\savebox{\DERIVBOXZLM}[2.5em]{$\Longrightarrow\hspace{-1.5em}
\raisebox{.2ex}{*}
\hspace{-.7em}\raisebox{-.8ex}{\scriptsize lm}\hspace{.7em}$}

\begin{document}
\bibliographystyle{apsrev}
\epsfclipon


\newcommand{\pbp}{\langle \bar \psi \psi \rangle}
\newcommand{\pbdmdup}{\left\langle \bar \psi \frac{dM}{du_0} \psi
\right\rangle}


\title{QCD thermodynamics with nonzero chemical potential at $N_t=6$
and
effects from heavy quarks}

\author{C.~DeTar and L.~Levkova} \affiliation{Physics Department, University of Utah,
Salt Lake City, Utah 84112, USA}

\author{Steven Gottlieb
\footnote{On sabbatical leave at NCSA, University of Illinois, Urbana IL 61801, USA}}
\affiliation{Department of Physics, Indiana University, Bloomington,
Indiana 47405, USA
}

\author{U.M.~Heller}
\affiliation{American Physical Society, One Research Road, 
Ridge, New York 11961, USA}

\author{J.E.~Hetrick}
\affiliation{Physics Department, University of the Pacific, Stockton, California 95211, USA}

\author{R.~Sugar}
\affiliation{Department of Physics, University of California, Santa
Barbara, California 93106, USA}

\author{D.~Toussaint}
\affiliation{Department of Physics, University of Arizona, Tucson, Arizona
85721, USA}
\date{\today}

\begin{abstract} 
We extend our work on QCD thermodynamics with 2+1 quark flavors at nonzero chemical
 potential to finer lattices with $N_t=6$. We study the equation of state and other thermodynamic 
quantities, such as quark number densities and susceptibilities, and compare them with 
our previous results at $N_t=4$. We also calculate the effects of the addition
of the charm and bottom quarks on the equation of state at zero and nonzero chemical potential.
These effects are important for cosmological studies of the early Universe.
\end{abstract}

\pacs{12.38.Gc, 12.38.Mh, 25.75.Nq}

\maketitle

\newpage


\section{Introduction}
\label{sec:intro}
The quark-gluon plasma (QGP) is a state of matter which
forms at very high temperatures or densities. It is believed 
that up to microseconds after the big bang 
the QGP was a dominant component of the Universe.
This state of matter is recreated
in heavy-ion collision experiments [such as are done at the Relativistic
Heavy Ion Collider (RHIC)] which study its
formation and transition to ordinary matter.
The equation of state (EOS) of the QGP is essential to our understanding of 
its hydrodynamic expansion and consequently of 
the particle 
spectra produced in these experiments. We have studied the EOS at zero
and nonzero chemical potential previously \cite{Bernard:2006nj,Bernard:2007nm}.
Here we extend our
work in two directions. (1) We present results for the EOS at nonzero chemical
potential at finer lattice spacings 
than our previous work.
Here the temporal lattice extent is $N_t=6$, where previously it was $N_t=4$. 
Preliminary results for the $N_t=6$ case were reported in
Ref.~\cite{Basak:2009uv}.
It is important to compare the two cases and determine the size of 
the discretization error as a step towards taking the continuum extrapolation.
(2) We include the effects of the charm and bottom quarks.
A preliminary progress report on the charm quark effects was given in Ref.~\cite{Levkova:2009gq}.
We use the heavy-quark-quenched approximation.  That is, the
charm and bottom quarks appear as valence quarks, but not as dynamical sea quarks.
Thus we ignore all
charm and bottom quark loops contributing to the operators we determine in order to obtain the EOS. This approximation
introduces an error in our calculation. However, considering that the charm
and bottom quarks are much heavier than our sea $u$, $d$, and $s$ quarks, it seems plausible that adding
sea charm and bottom quarks would have a small effect for temperatures
much less than their masses. Still, until we have
a dynamical $c$- and $b$-quark calculation to compare against, this statement remains a conjecture.
The equation of state with the charm and bottom quarks added is most
applicable to the study of the early Universe, since the time scale relevant 
to the heavy-ion collisions at RHIC is probably 
too short for the charm and bottom quarks to thermalize and have a visible effect on the particle data.

As in our previous $N_t=4$ determination of the EOS
at nonzero chemical potential, we employ the Taylor expansion method.
For a detailed description of the method, see Refs.~\cite{Allton:2002zi,Gavai:2003mf}. The expansion
is carried up to sixth order in the expansion parameters $\mu_q/T$, where $\mu_q$
is the chemical potential for a certain quark flavor $q$ and $T$ is the temperature.

The gauge ensembles we used in this work are
the same as in Ref.~\cite{Bernard:2007nm}. They are generated using the asqtad 
improved staggered action \cite{Orginos:1998ue} and
have two degenerate light quarks and a strange quark in the sea. The ensembles lie
approximately on a trajectory of constant physics, where the strange 
quark mass $m_s$ is tuned to be close to its physical value,
and the light quark mass $m_l$ is one-tenth of $m_s$. Because in this paper we also consider charm and bottom quarks,
we do not refer to the strange quark as the ``heavy quark''
as in Refs. \cite{Bernard:2006nj,Bernard:2007nm}. 

In Sec.~II, we present our results for the 2+1 flavor EOS with nonzero chemical potential
at $N_t=6$, and 
compare it with our previous one at $N_t=4$. We also show other thermodynamic quantities,
such as the quark number susceptibilities and light-quark density.
Section~III gives our findings for the isentropic EOS
for 2+1 flavors. In Sec.~IV, we calculate the effects of the charm quark on the EOS
at zero and nonzero chemical potential, using the 
heavy-quark-quenched 
approximation to represent it. 
Section~V does the same for the bottom quark. In Sec.~VI, we give our conclusions.
The Appendixes contains some helpful formulas for the application
of the Taylor expansion method for the EOS calculation in the 2+1+1 quark flavor case.
 
\section{The EOS at nonzero chemical potential at $N_t=6$ for 2+1 flavors}
The Taylor expansion method allows us to represent the pressure $p$ and the interaction 
measure $I$ in the case where both the light and the strange quark chemical potentials are nonzero,
as the following infinite sums:
\bea
{p\over T^4}&=&{\ln {\cal Z} \over T^3V}=
\sum_{n,m=0}^\infty c_{nm}(T) \left({\bar{\mu}_l\over T}\right)^n
\left({\bar{\mu}_s\over T}\right)^m,
\label{p}\\
{I\over T^4}&=&-{N_t^3\over N_s^3}{d\ln {\cal Z} \over d\ln a}=\sum_{n,m}^\infty b_{nm}(T)\left({\bar{\mu_l}\over T}\right)^n
\left({\bar{\mu_s}\over T}\right)^m.
\label{I}
\eea
 In the above, $\bar{\mu}_{l,s}$ are the chemical potentials for the 
light and strange quarks in physical units, $T$ is the temperature, $N_s$ is the spatial lattice extent,
${\cal Z}$ is the partition function and $a$ is the lattice spacing.
\begin{figure}[t]
  \epsfxsize=16cm
  \epsfbox{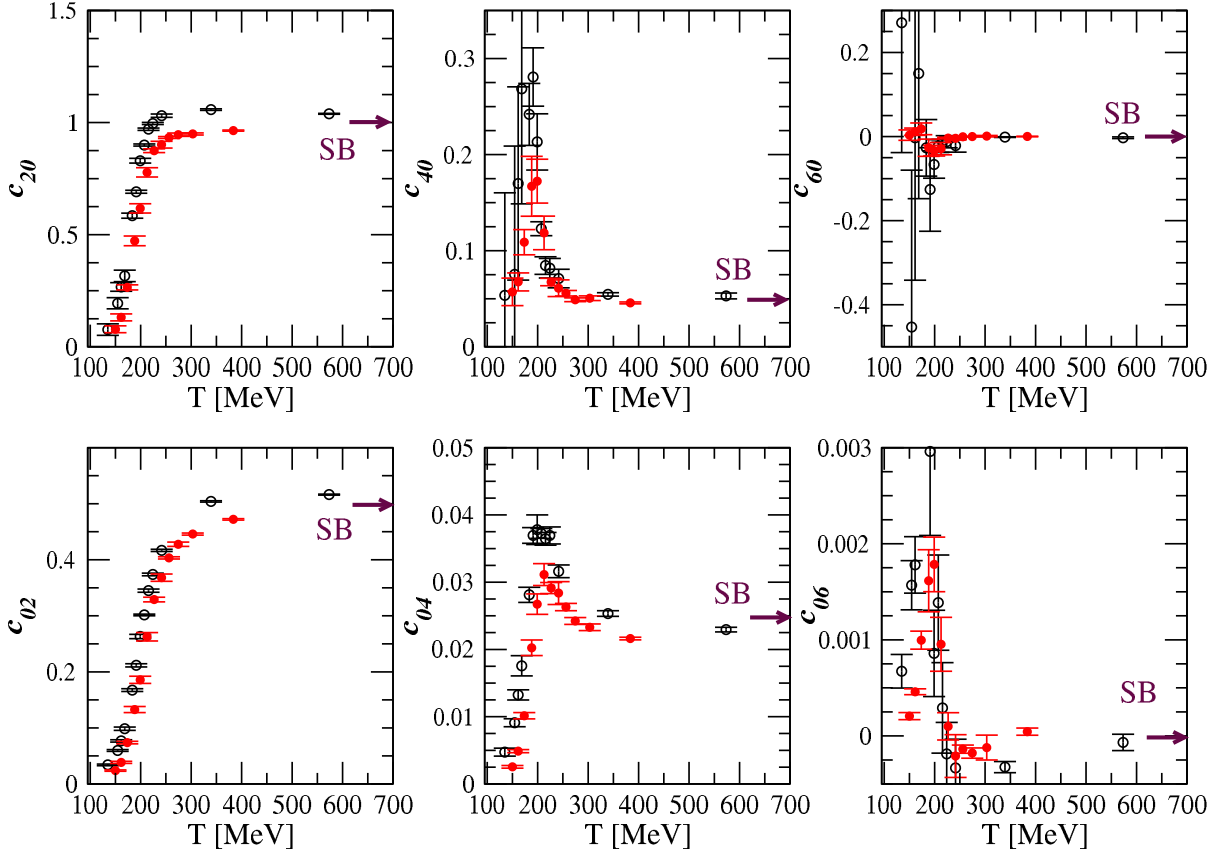}
\caption{Unmixed coefficients $c_{n0}$ and $c_{0n}$ in the Taylor expansion of 
the  pressure
as a function of temperature. The new results for
 $N_t=6$ are shown in filled (red) circles; empty (black) circles 
are used for $N_t=4$ (from Ref.~\cite{Bernard:2007nm}). Arrows indicate the
Stefan-Boltzmann limit for each of the coefficients.
}
\label{fig:C_unmixed}
\end{figure}
Because of $CP$ symmetry, the expansion coefficients $c_{nm}$ and $b_{nm}$ are nonzero 
only if $n+m$ is an even integer. The explicit forms of these coefficients are
given in Ref.~\cite{Bernard:2007nm}, and since they are somewhat involved, we do not repeat them here. 
We calculate the coefficients stochastically with random Gaussian sources. 
Inside the transition region, we used 800 random sources per lattice, and outside, 400.
With these numbers, the stochastic error in the unmixed second order
coefficients, {\it i.e.}, the diagonal quark number susceptibilities at zero chemical potential, 
is about 20\%
of the full statistical error.
These coefficients are the ones with largest contribution to the thermodynamic 
quantities for each type of quark.
For the fourth order unmixed coefficients, this error is
about 50\% of the final error. For the rest of the coefficients (mixed coefficients and all
coefficients of sixth order), the contribution of the 
stochastic error is dominant. A further increase of the number of sources
as a way to decrease the stochastic noise seems impractical at this point.  
We need either significantly more computer power or a substantial improvement of the 
noisy estimators in order to reduce the resulting stochastic error. 

\begin{figure}[t]
  \epsfxsize=16cm
  \epsfbox{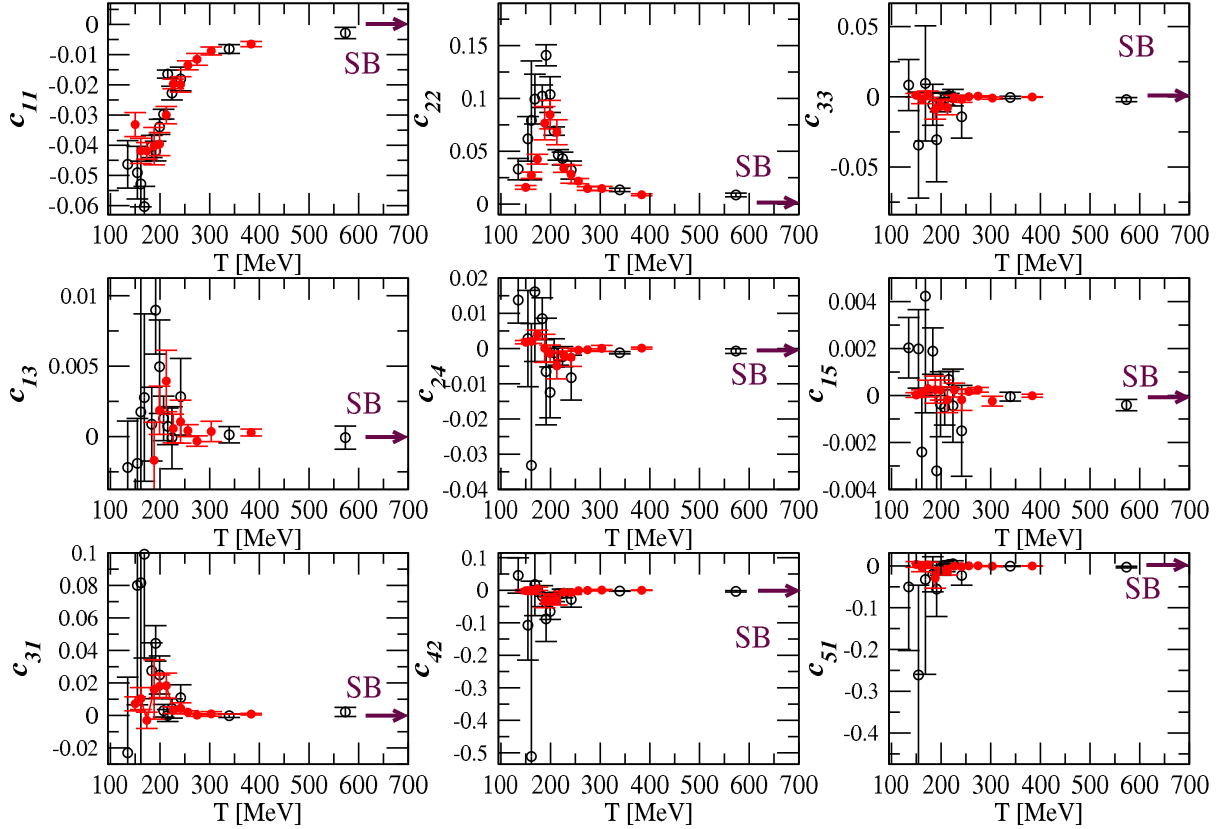}
\caption{Mixed coefficients $c_{nm}$  in the Taylor expansion of 
the pressure 
as a function of temperature. The new results for
 $N_t=6$ are shown in filled (red) circles; empty (black) 
circles are used for $N_t=4$ (from Ref.~\cite{Bernard:2007nm}).
Arrows indicate the
Stefan-Boltzmann limit for each of the coefficients.
}
\label{fig:C_mixed}
\end{figure}

\begin{figure}[t]
  \epsfxsize=16cm
  \epsfbox{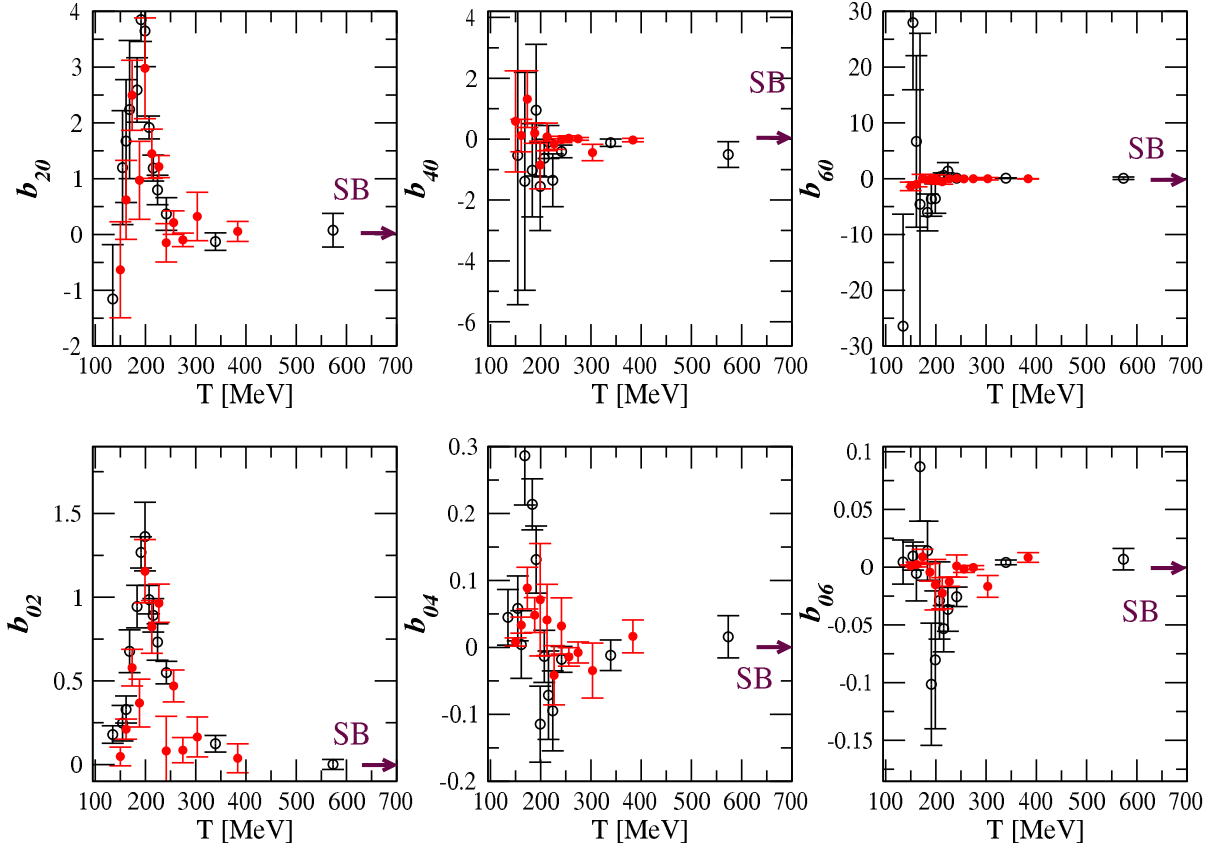}
\caption{Unmixed coefficients $b_{n0}$ and $b_{0n}$ in the Taylor expansion of 
the interaction measure 
as a function of temperature. The new results for
 $N_t=6$ are shown in filled (red) circles; empty (black) circles 
are used for $N_t=4$ (from Ref.~\cite{Bernard:2007nm}).
Arrows indicate the
Stefan-Boltzmann limit for each of the coefficients.
}
\label{fig:B_unmixed}
\end{figure}

\begin{figure}[t]
  \epsfxsize=16cm
  \epsfbox{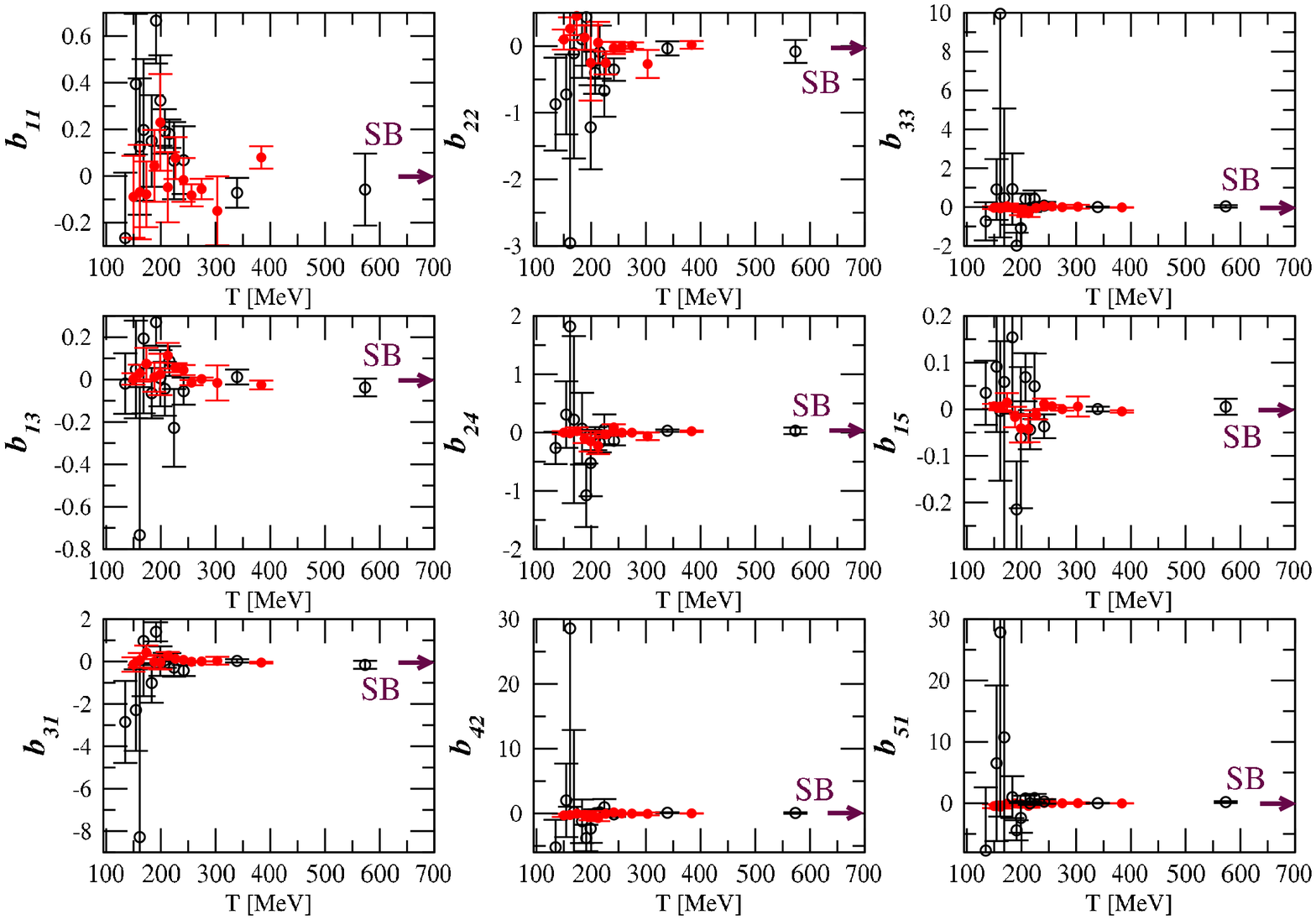}
\caption{Mixed coefficients $b_{mn}$ in the Taylor expansion of 
the interaction measure as a function of temperature. 
New results for $N_t = 6$ are shown in
filled (red) circles, empty (black) circles are used for $N_t = 4$ (from Ref.~\cite{Bernard:2007nm}).
Arrows indicate the
Stefan-Boltzmann limit for each of the coefficients.
}
\label{fig:B_mixed}
\end{figure}

Figures~\ref{fig:C_unmixed} and \ref{fig:C_mixed} show some of the coefficients  
in the pressure expansion and compare our new results at $N_t=6$ (red filled circles)
 with the previous ones \cite{Bernard:2007nm}
at $N_t=4$~(black empty circles).
We can see that the errors for the $N_t=6$ case are smaller than the ones at the shorter temporal extent,
due to both the increased volume and increased number of random sources.
(We previously used 100--200 sources.)
There is also a shift in the central values between the two cases which indicates that
the discretization effects at $N_t=4$ are significant. The approach of the coefficients
to the (massless) Stefan-Boltzmann
continuum limit with increasing $T$ in the case of $N_t=6$ is  slower, and the structure at low temperature is 
made somewhat clearer due to the smaller errors on the data. 
Similarly, Figs.~\ref{fig:B_unmixed} and \ref{fig:B_mixed} compare the unmixed 
and mixed coefficients $b_{nm}$ involved in the Taylor expansion of the interaction measure
at the two different temporal extents. These coefficients are calculated independently
from the ones in the pressure expansion (although the two sets of coefficients are 
technically related by integration). The operators involved in the determination
of the interaction measure expansion coefficients
are intrinsically noisier, which is reflected in the larger errors on the data. 
Still, the shift in central values and errors in the $b_{nm}$
coefficients in going from $N_t=4$ to $6$ is qualitatively similar to that of
the pressure coefficients $c_{nm}$.

\begin{figure}[t]
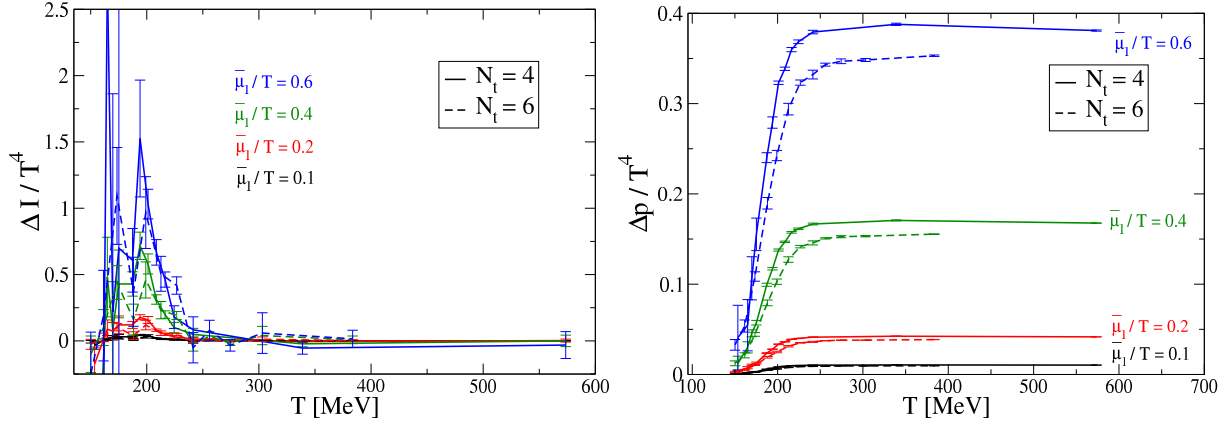

  \includegraphics*[width=8cm]{dI.eps}
  \includegraphics*[width=8cm]{dP.eps}
\caption{(Left panel) The change in the interaction measure due to the nonzero chemical potentials {\it vs.}
temperature. At a given $N_t$, the larger $\bar{\mu}_l/T$, the higher the data appears
on the plot. (Right panel) Similarly, the change in the pressure.
}
\label{fig:dIdP}
\end{figure}

\begin{figure}[t]
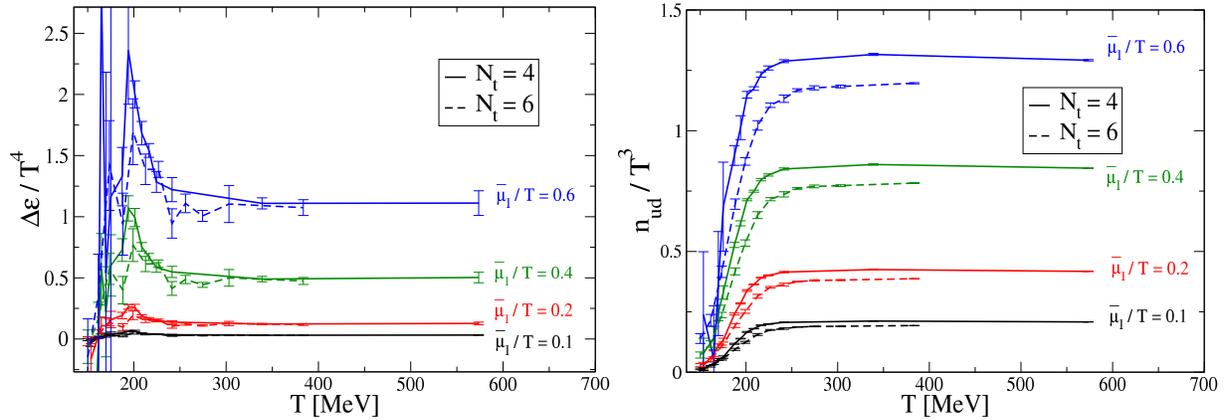

  \includegraphics*[width=8cm]{dE.eps}
  \includegraphics*[width=8cm]{dNud.eps}
\caption{(Left panel) The change in the energy density due to the nonzero chemical potentials {\it vs.}
temperature. (Right panel) The light-quark density {\it vs.} temperature for several values of
$\bar\mu_l/T$ (in different colors) and $\mu_s$ tuned such that $n_s\approx 0$.
}
\label{fig:dE_Nl}
\end{figure}

\begin{figure}[t]
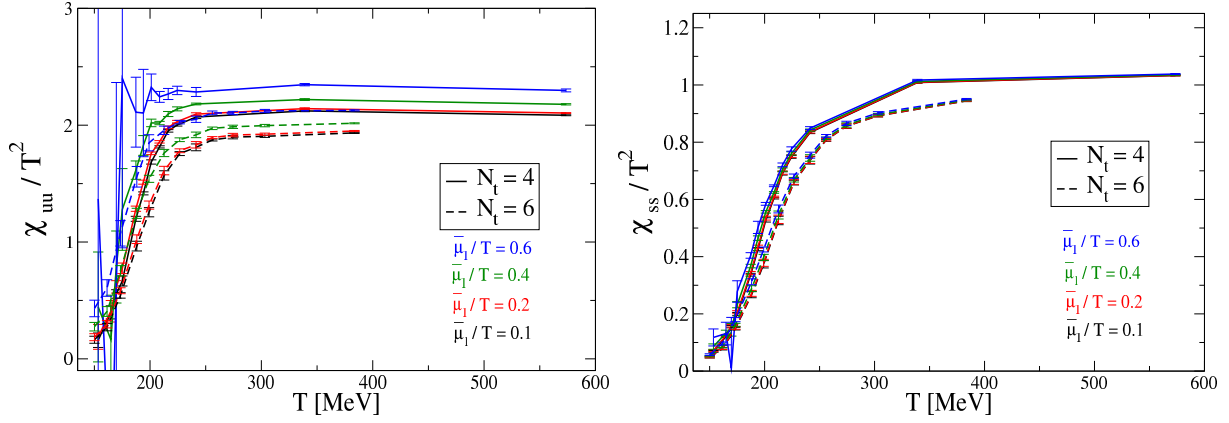

  \includegraphics*[width=8cm]{dXuu.eps}
  \includegraphics*[width=8cm]{dXss.eps}
\caption{(Left panel) The light-light quark number susceptibility {\it vs.} temperature for several values of
$\bar\mu_l/T$ (in different colors) and $\mu_s$ tuned such that $n_s\approx 0$. At a given $N_t$, the larger $\bar{\mu}_l/T$, the higher the data appears
on the plot. However, at low $\bar{\mu}_l/T$ these differences are very small. (Right panel)
Same for the strange-strange quark number susceptibility. Here the data dependence on $\bar{\mu}_l/T$ is 
very weak.
}
\label{fig:Xuu_Xss}
\end{figure}

\begin{figure}[t]
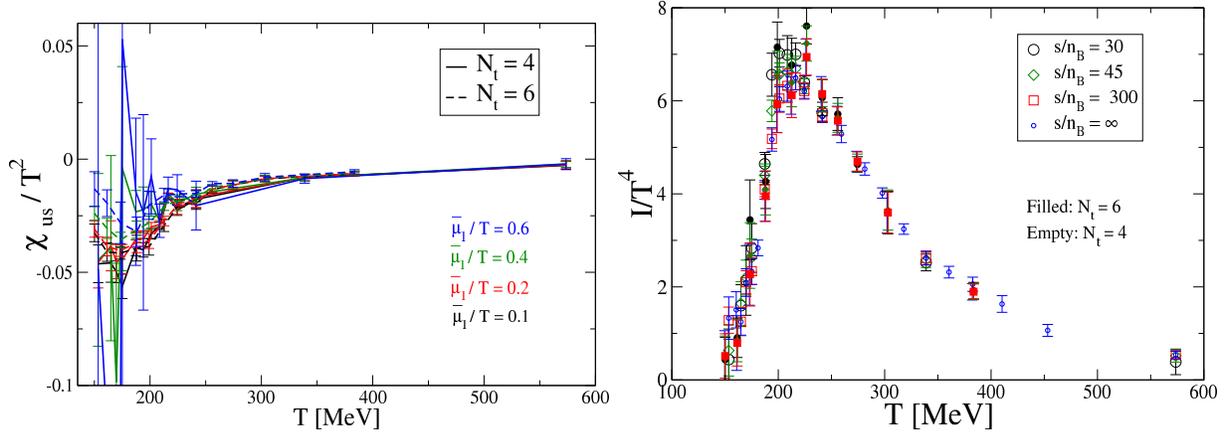

  \includegraphics*[width=8cm]{dXus.eps}
  \includegraphics*[width=8cm]{IO6Isentr.eps}
\caption{(Left panel) The light-strange quark number susceptibility {\it vs.} temperature for several values of
$\bar\mu_l/T$ (in different colors) and $\mu_s$ tuned such that $n_s\approx 0$. As in Fig.~\ref{fig:Xuu_Xss} (right panel),
the data dependence on $\bar{\mu}_l/T$ is 
very weak.
(Right panel) The isentropic interaction measure {\it vs.} temperature for selected values
of $s/n_B$. 
}
\label{fig:Xus_Iisentr}
\end{figure}

Having obtained the coefficients in the Taylor expansions in Eqs.~(\ref{p})~and~(\ref{I}), we can now
turn to 
calculating the EOS. Because of 
the nonzero $c_{n1}(T)$ terms, a nonzero strange quark density $n_s$ is 
induced even with $\mu_s = 0$.
(We use $\mu_f$ to denote the chemical potential in lattice units for flavor $f$). To study the
$n_s = 0$ plasma, we must, therefore, 
tune $\mu_s$ as a function of $\mu_l$ and $T$. Figures~\ref{fig:dIdP} (both panels)~and~\ref{fig:dE_Nl}~(left panel)
show the changes
in the interaction measure ($\Delta I$), pressure ($\Delta p$) and energy density ($\Delta \varepsilon=\Delta I+3\Delta p$)
at $\bar\mu_l/T = 0.1,\,0.2,\,0.4$, and $0.6$, and with $\bar\mu_s/T$ tuned along the trajectory so that $n_s\approx 0$.
We see statistically significant discretization effects when we compare the $N_t=4$ and 6 
cases for $\Delta p$, with the latter data lying lower than the former. 
The $\Delta I$ results have larger errors, and discerning differences in the data from the two
temporal extents is more difficult. The $N_t=6$ data is slightly but consistently 
lower than the one from the $N_t=4$ calculation; however, this is
not a statistically significant observation. The change in the energy density $\Delta \varepsilon$
inherits the large errors from $\Delta I$, and the same conclusions apply for it.
The discretization effects for the light-quark density ($n_{ud}$), the light-light and strange-strange quark number
susceptibilities ($\chi_{uu}$ and $\chi_{ss}$, respectively), are examined in Figs.~\ref{fig:dE_Nl}~(right panel)~and~\ref{fig:Xuu_Xss}~(both panels). The
effect of increasing the temporal extent from $N_t=4$ to 6 is to lower these quantities by 4\%--10\%.
For the range of values of $\bar\mu_l/T$ that we examine, we do not find any evidence for peaks that could presage critical behavior in  $\chi_{uu}$.
The light-strange quark number susceptibility ($\chi_{us}$), shown in Fig.~\ref{fig:Xus_Iisentr} (left panel),
is too noisy for
a reliable conclusion about its discretization effects; there is only a hint at a possible move
toward lower absolute values at the larger $N_t$.
\section{The isentropic equation of state}

\begin{figure}[t]
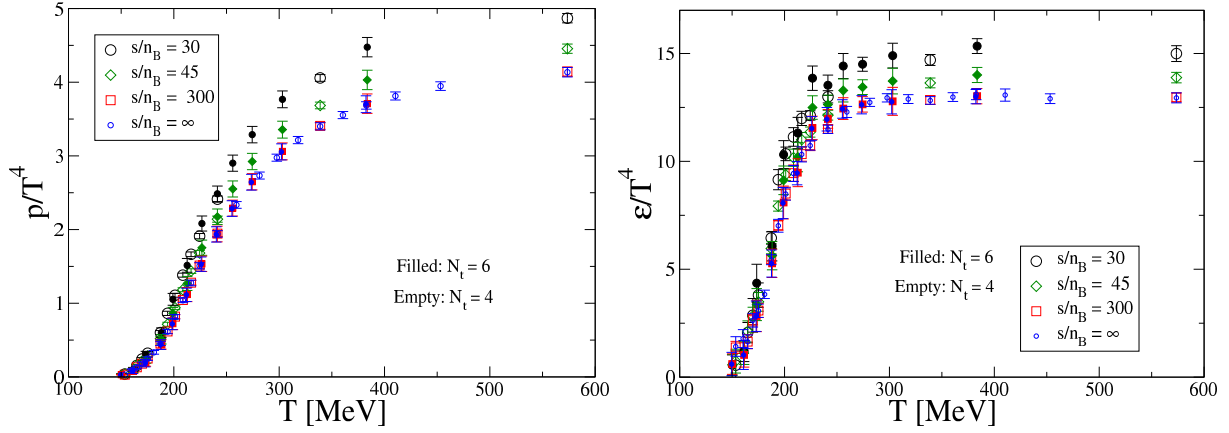

 \includegraphics*[width=8cm]{PO6isentr.eps}
 \includegraphics*[width=8cm]{EO6Isentr.eps}
\caption{(Left panel) The isentropic pressure {\it vs.} temperature for selected values
of $s/n_B$. (Right panel) The same for the isentropic energy density.
}
\label{fig:PEisentr}
\end{figure}

\begin{figure}[t]
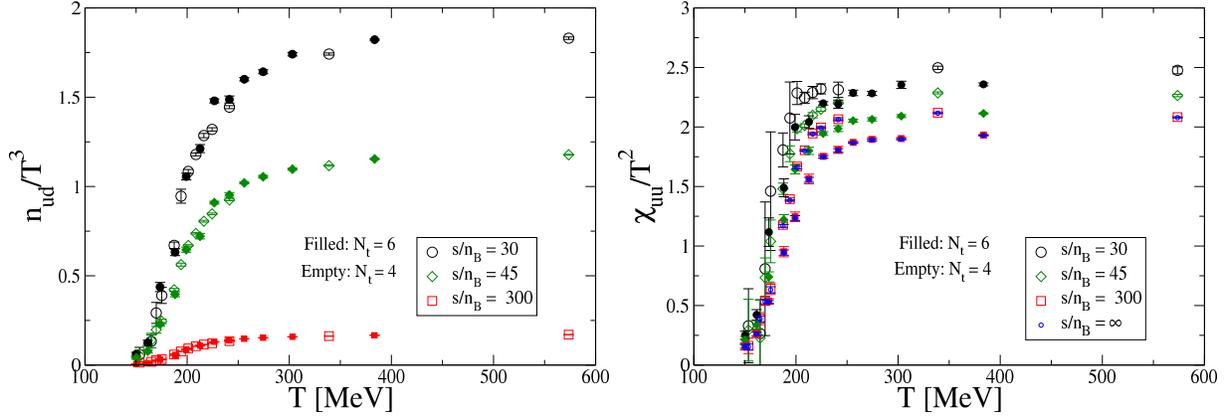

 \includegraphics*[width=8cm]{nuO6Isentr.eps}
 \includegraphics*[width=8cm]{Chi_uuO6Isentr.eps}
\caption{(Left panel) The isentropic light-quark density {\it vs.} temperature for selected values
of $s/n_B$. (Right panel) The same for the isentropic light-light quark number susceptibility.
}
\label{fig:Xuu_Nuisentr}
\end{figure}

\begin{figure}[t]
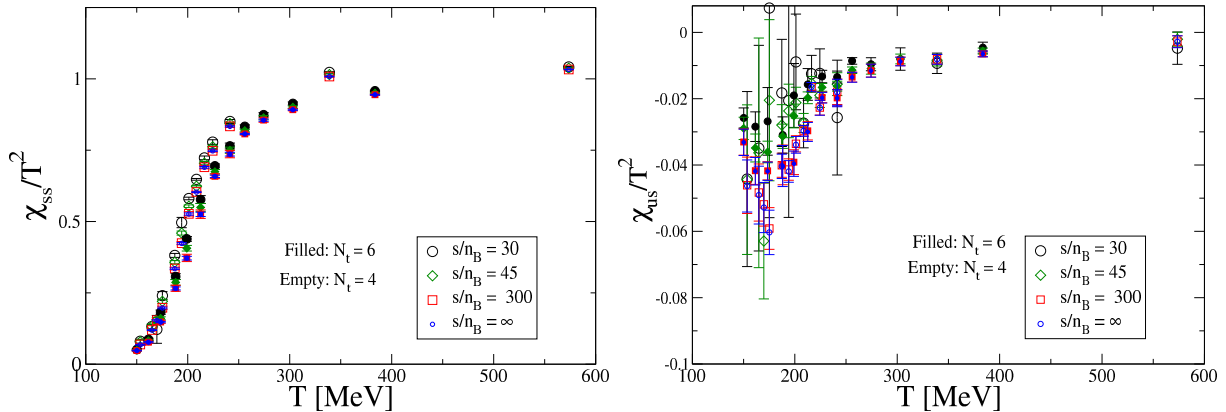

 \includegraphics*[width=8cm]{Chi_ssO6Isentr.eps}
 \includegraphics*[width=8cm]{Chi_usO6Isentr.eps}
\caption{(Left panel) The isentropic strange-strange quark number susceptibility
 {\it vs.} temperature for selected values
of $s/n_B$. (Right panel) The same for the isentropic light-strange quark number susceptibility. 
}
\label{fig:Xssisentr}
\end{figure}

The form of the EOS most applicable to the experimental conditions of the heavy-ion collisions
is the isentropic
one. There, after thermalization, the system expands and cools with 
constant entropy. We determine the isentropic EOS by performing our calculations at 
fixed ratio of entropy to baryon number ($s/n_B$). This is achieved by finding the trajectories
in the ($\mu_l$, $\mu_s$, $T$) space which satisfy (within errors)
both $s/n_B=C$ and $n_s=0$, where $C$ is
a constant whose value depends on the particular experiment we are interested in.
For AGS, SPS, and RHIC, we have $s/n_B=30$, 45, and 300, respectively. 
Figures~\ref{fig:Xus_Iisentr} (right panel)~and~\ref{fig:PEisentr}~(both panels)
show our results for the interaction measure, pressure, and energy density for the
different $s/n_B$ values appropriate for these experiments. We compare the new results
at $N_t=6$ (filled symbols) with the ones already published in Ref.~\cite{Bernard:2007nm} at $N_t=4$ (empty symbols).
 For these three quantities, the comparison shows negligible effects due to the increase 
of the temporal extent $N_t$. The reason for this is that by far the largest contribution 
to these quantities---the zero-chemical potential (zeroth order) term in their respective
Taylor expansions---
does not show 
large discretization effects \cite{Bernard:2006nj}. On the other hand, quantities which do not
have a zeroth order term may show larger differences between the $N_t=6$ and 4 cases.
Indeed, small discretization effects are evident in the isentropic 
light-light and strange-strange quark number
susceptibilities in Figs.~\ref{fig:Xuu_Nuisentr} (right panel)~and~\ref{fig:Xssisentr} (left panel),
respectively. However, the isentropic light-quark density, shown in Fig.~\ref{fig:Xuu_Nuisentr} (left panel),
has only marginal discretization effects, despite the fact that it does not have a
zeroth order contribution. The large errors on the strange-light susceptibility,
shown in Fig.~\ref{fig:Xssisentr} (right panel),
precludes us from drawing a conclusion about its discretization effect.    
To conclude with our final observation, for both values of $N_t$, we find
rather smooth behavior for the isentropic variables indicating that experiments 
are far from any critical point in the $\mu-T$ plane.
\section{The effects of the charm quark on the EOS}
In this section, we study the effects of the charm
quark on the EOS at zero and nonzero chemical potential. Our preliminary results were reported 
in Ref.~\cite{Levkova:2009gq}.
First, let us discuss the relevance of the charm quark contribution.
The experiments at RHIC create a ``fireball'' which thermalizes within $\tau\approx10^{-24}$ s \cite{Adcox:2004mh}.
The $u$, $d$, and $s$ quarks participate in the thermal ensemble describing the state of the thermalized
fireball. Under the experimental conditions, the $c$ quark probably
is not thermalized, and thus  the 2+1 flavor EOS is considered sufficient
for the hydrodynamics models applied to the current experimental data. The question of equilibration of charm,
however, is not completely settled as argued, for example, in
Ref.~\cite{Laine:2009ik}.
Furthermore, the situation may change for
the future LHC experiments.
A quark-gluon plasma also existed microseconds after the big bang. Under these primordial conditions and longer time scales,
the $c$ quark probably participated in the thermal ensemble as well, which implies that for 
the study of the early Universe, the EOS with 2+1+1 flavors would be important \cite{Laine:2006cp}. For example,
the scale factor of the early Universe is affected by the number 
of quark flavors in the EOS used for its determination \cite{McGuigan:2008pz}. 
Previously, the question of the charm quark contribution to the EOS at zero chemical potential
has been studied on the lattice in Ref.~\cite{Cheng:2007wu} at $N_t=4$, 6, and 8
using the p4 fermion formulation.
That study treated the charm quark as a valence staggered quark. We do the
same, but in the asqtad formulation at $N_t = 6$. 
We tuned the charm quark using a different strategy
than in Ref.~\cite{Cheng:2007wu}, where the charm quark mass was determined using the
$\eta_c$ or $J/\Psi$ rest mass on all available ensembles.
In our study, the charm quark mass was tuned to match the rest mass of the $D_s$ at $\beta=7.08$ 
($a\approx0.086$~fm)
where the discretization effects are smallest on our trajectory. We chose the
$D_s$ for our tuning purposes because the discretization effects are smaller
for heavy-light mesons than for the heavy-heavy ones~\cite{Kronfeld:1996uy}. 
We found $m_c/m_s=10$ at our tuning point with a 4\% uncertainty.
We have kept this ratio constant for lower temperatures. It is probably 
incorrect at the lowest 
 available temperatures, but due to large discretization effects, the tuning 
is inherently problematic there. Still, we do not expect this to matter much,
thanks to the large mass of the $c$ quark and 
its very small contribution in that region. 

Following our method in Ref.~\cite{Bernard:2006nj} for determining the EOS
at zero chemical potential, the  
2+1+1 flavor interaction measure was obtained by adding to
our previous results for the 2+1 flavor case the charm contribution
\be
\label{eq:I}
I_ca^4= -\frac{1}{4}\left[\frac{d (m_c a)}{d \ln a}\Delta\VEV{\bar{\psi} \psi}_c
+ \frac{d u_0}{d \ln a}\Delta\VEV{\bar{\psi}\,\frac{d M}{d u_0}\,\psi}_c\right],
\ee
where the observables in the above are calculated in the 
heavy-quark-quenched 
approximation and $\Delta$
stands for the difference between the zero and nonzero temperature value of an observable. 
The mass beta function is
approximated as 
\be
\frac{d (m_c a)}{d \ln a} = 10\frac{d (m_s a)}{d \ln a},
\ee
since we kept the ratio $m_c/m_s=10$ constant along the trajectory. We determined the strange quark mass beta function 
and the function $d u_0/d \ln a$ previously \cite{Bernard:2006nj}.
To find the
charm contribution to the pressure and energy density, we integrated Eq.~(\ref{eq:I})
along the physics trajectory, as in Ref.~\cite{Bernard:2006nj} for the 2+1
flavor case.

Again, the nonzero chemical potential calculation was done using the Taylor expansion method, taken
to sixth order. 
For 2+1+1 quark flavors, the Taylor expansion of the pressure is modified to the following form:
\be
{p\over T^4}=
\sum_{n,m,k=0}^\infty c_{nmk}(T) \left({\bar{\mu}_l\over T}\right)^n
\left({\bar{\mu}_s\over T}\right)^m\left({\bar{\mu}_c\over T}\right)^k,
\ee
where $\bar{\mu}_{l,s,c}$ are the chemical potentials in physical units for
the light ($u,d$), strange ($s$)
and charm ($c$) quarks.
Because of $CP$ symmetry the terms in the above are nonzero only if $n+m+k$ is even.
The interaction measure has the same form with only $c_{nmk}\rightarrow b_{nmk}$. 
Some details of the explicit calculations for the  
pressure and interaction measure coefficients can be found in the Appendixes.
\begin{figure}[t]
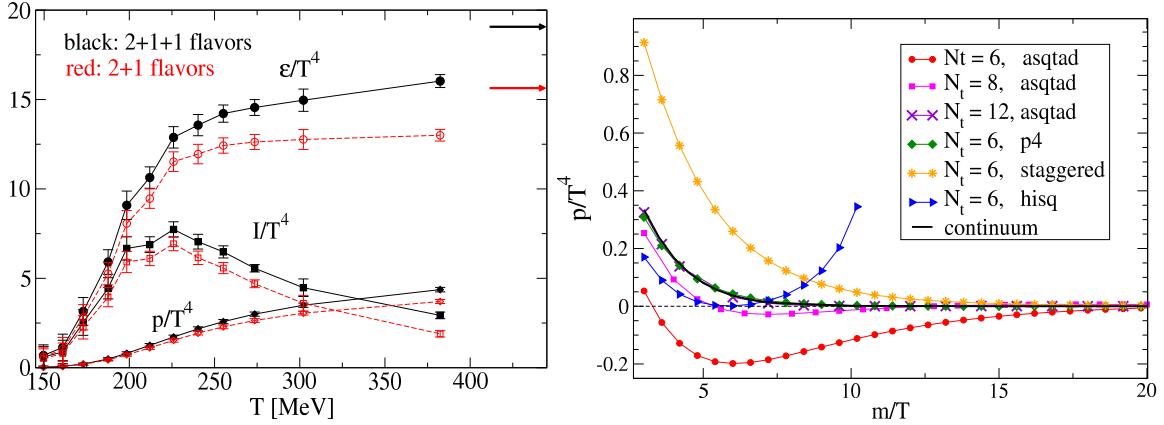

 \includegraphics*[width=7.2cm]{EOS_charm.eps}
 \includegraphics*[width=8cm]{Free_pressure.eps}
\caption{(Left panel) Interaction measure ($I$), pressure ($p$) and energy density ($\varepsilon$ ) 
divided by the temperature to the
fourth power ($T^4$) for the cases of 2+1 (red) and 2+1+1 (black) flavors. The arrows indicate
the  energy density Stefan-Boltzmann limit for both cases. (Right panel) The pressure for 1 quark flavor in the free theory
{\it vs.} the ratio of the quark mass ($m$) and temperature ($T$) for different staggered quark formulations.
The rise of the pressure at large $m/T$ in the HISQ case shows that higher order corrections to the
Naik term are needed in this region. Currently we have corrections up to $O(m^8)$ only.
}
\label{fig:eos_charm}
\end{figure}
We used 800 random sources per lattice
in the transition region and 400 outside it to calculate the new observables in the expansions of
the pressure and interaction measure.
For the calculation at nonzero chemical potential, the valence $c$ quark had a low cost 
in terms of computer time, but it required
a sizable software development. 
For $2+1$ flavors we had $95$ observables to code and for  $2+1+1$
flavors there were $399$.

Turning to our results, let us first examine 
the effects of the charm
quark on the EOS at zero chemical potential. Figure~\ref{fig:eos_charm} (left panel) 
shows our results for the EOS with 2+1+1 flavors
and compares it with previous results for 2+1 flavors
\cite{Bernard:2006nj}. The charm 
quark contribution grows with temperature, as expected, and at the highest
available $T$ it contributes about 20\% to the energy density. We conclude that
in the cases where the charm quark is thermalized, its contribution  
to the EOS at temperatures
higher than about 200 MeV, cannot be ignored. Our result at $N_t=6$ is qualitatively similar to the
previous work \cite{Cheng:2007wu}, but quantitatively our charm quark contributions to
the energy density and pressure
are about 25\%-30\% lower  by comparison at temperatures around 400 MeV. A possible explanation for this
is the larger discretization effects for the heavy-quark pressure for the asqtad action than
for the p4 action. Figure~\ref{fig:eos_charm} (right panel) shows the free quark pressure as a function of the
ratio of the (heavy) quark mass and the temperature for different staggered lattice fermion formulations.
The asqtad action at $N_t=6$ shows a negative value for the pressure for a range of heavy-quark masses while the p4
action is close to the continuum limit. Our results for the charm contribution to the EOS 
do not show the outright unphysical behavior occurring in the free quark case, but it is possible that
the heavy-quark discretization effects
depress the lattice values.

\begin{figure}[t]
  \includegraphics*[width=16cm]{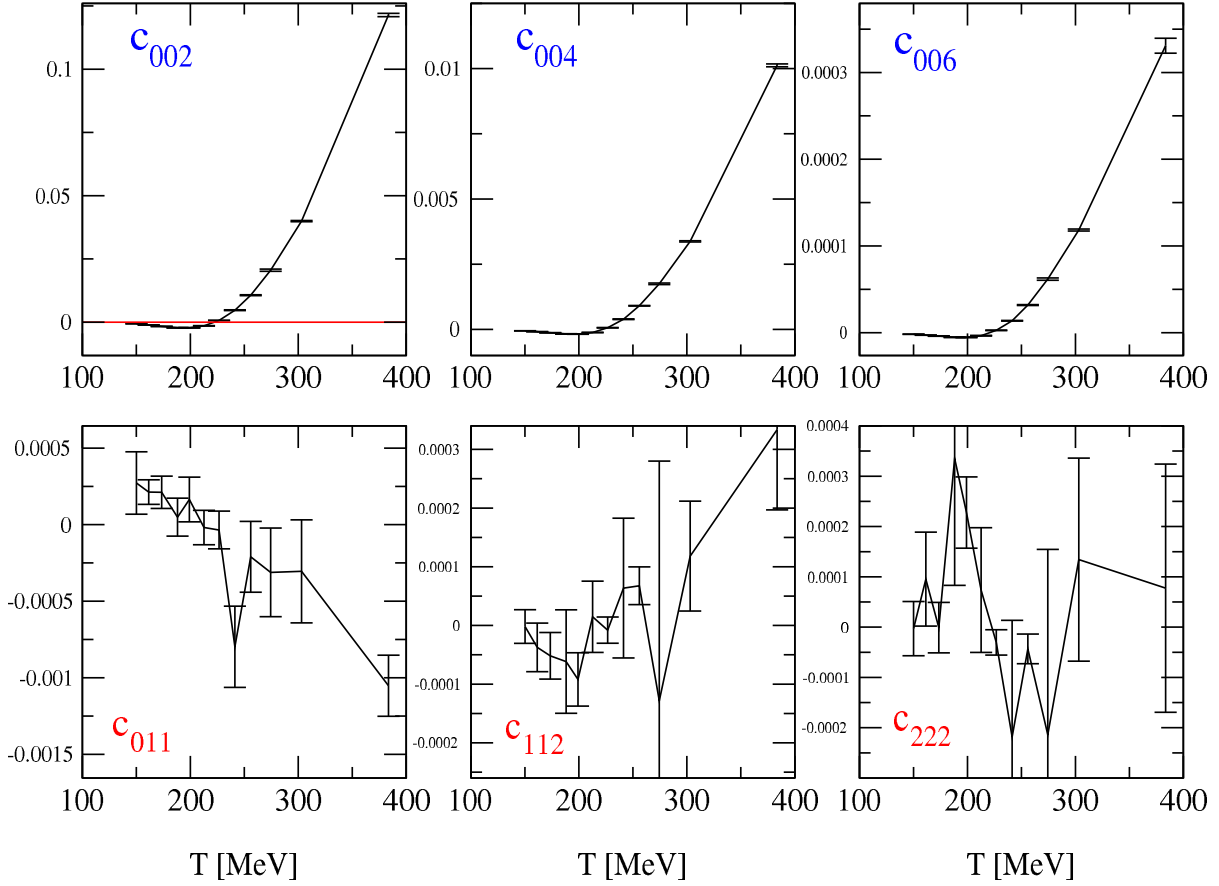}
\caption{Some of the new Taylor expansion coefficients for the pressure at nonzero chemical potential
when the charm quark is added to the partition function.
}
\label{fig:Pcoeff_charm}
\end{figure}

\begin{figure}[t]
  \includegraphics*[width=16cm]{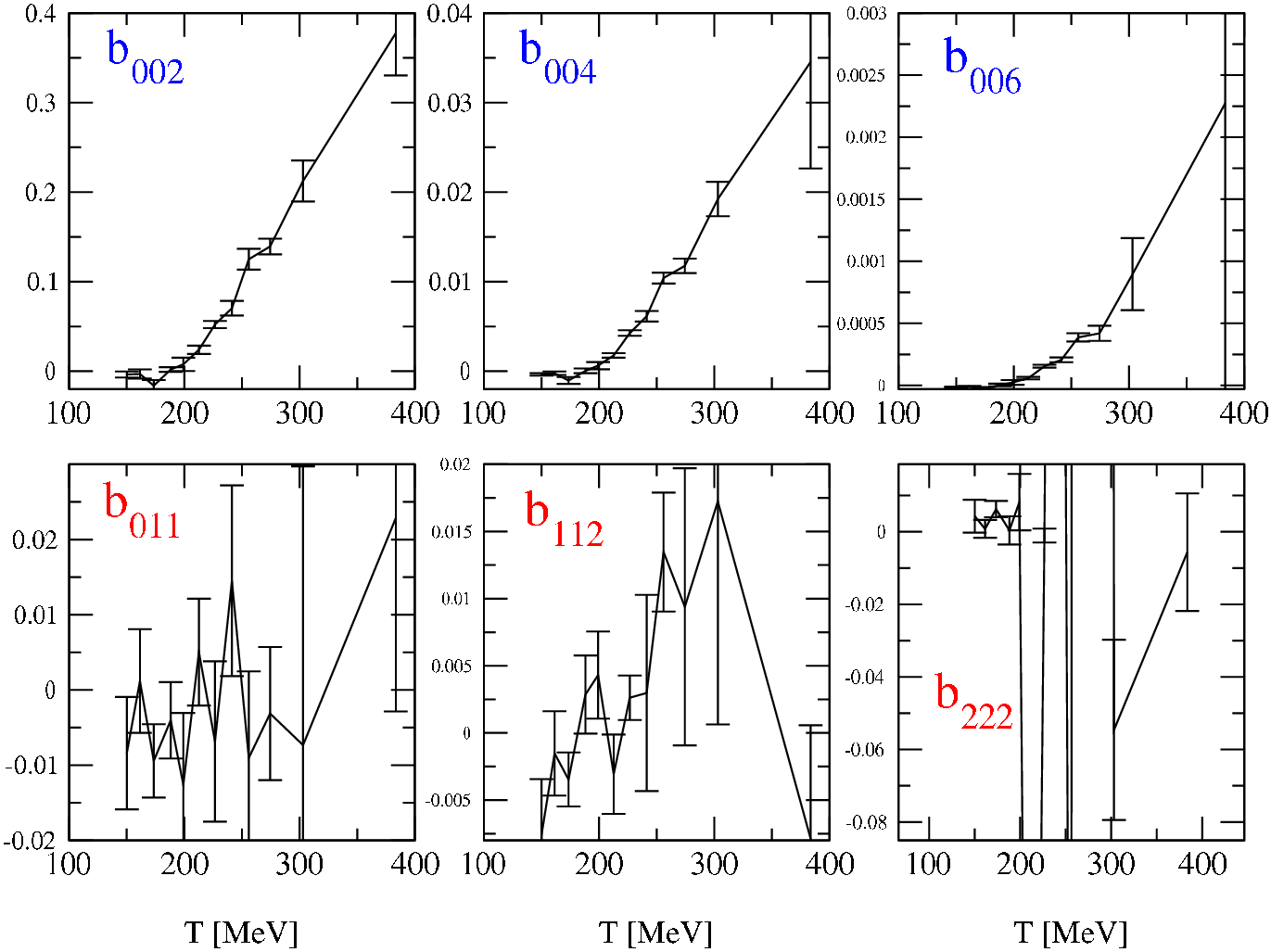}
\caption{Some of the new Taylor expansion coefficients for the 
interaction measure at nonzero chemical potential when the charm quark is added to 
the partition function.
}
\label{fig:Icoeff_charm}
\end{figure}

\begin{figure}[t]
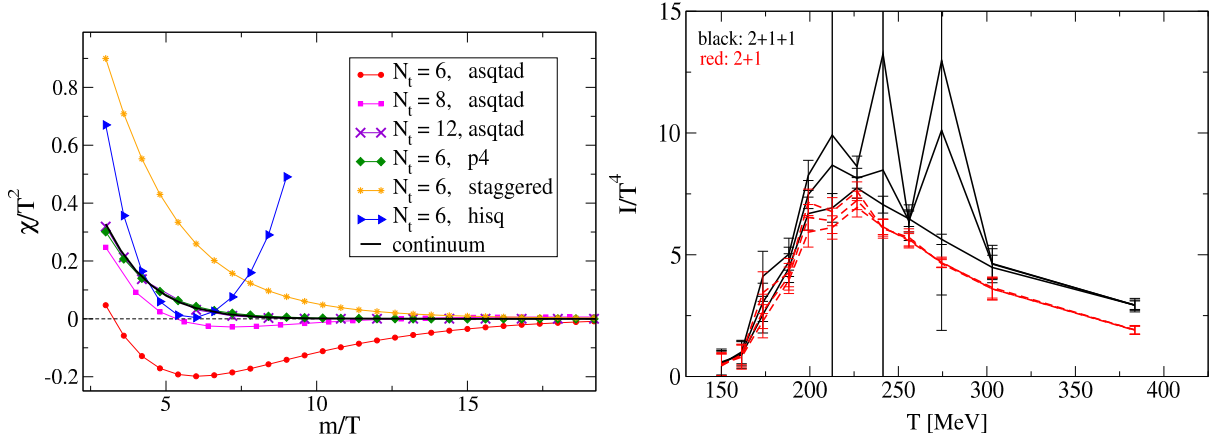

  \includegraphics*[width=7.9cm]{Chi_free_asqtad.eps}
  \includegraphics*[width=8cm]{I_isentr.eps}
\caption{(Left panel) The quark number susceptibility for 1 quark flavor in the free theory
{\it vs.} the ratio of the quark mass ($m$) and temperature ($T$) for different staggered quark formulations.
The reason for the rise of the susceptibility in the HISQ case for large $m/T$ is the same as explained in the caption of 
the right panel of Fig.~\ref{fig:eos_charm}.
(Right panel)  
The isentropic interaction measure at selected $s/n_B$ values 
for 2+1 and 2+1+1 flavors (red and black respectively). For a data set with the same color ({\it i.e.}, produced
with the same number of quark flavors), 
 the highest lying results are for
$s/n_B=30$, in the middle is the $s/n_B=45$ case and the case of $s/n_B=300$ has the lowest lying values.
}
\label{fig:free_Iisentr_charm}
\end{figure}

Now let us turn to the results at nonzero chemical potential. 
Figures~\ref{fig:Pcoeff_charm}~and~\ref{fig:Icoeff_charm} present some of the pressure and interaction
measure
expansion coefficients which are directly related to the charm quark contribution
at nonzero chemical potential. The first row in both figures shows the unmixed coefficients
and the second row---three of the mixed coefficients. 
The mixed 
coefficients are quite small and are much noisier than the unmixed ones, 
which was expected.
As a whole, the new unmixed coefficients 
$c_{00n}$ and $b_{00n}$ in 
the pressure and interaction measure expansions are small  compared with
the $c_{n00}$, $c_{0n0}$, $b_{n00}$, and $b_{0n0}$ coefficients. For numerical comparisons see Sec.~II,
where the latter four sets are defined
without the last zero in the subscripts. These new coefficients remain well below the 
continuum (massless) Stefan-Boltzmann values at the highest temperature available here. This is not 
surprising, since
over our temperature range $T < 2 T_c$, the charm quark mass is much
larger than the temperature.
The first panel of Fig.~\ref{fig:Pcoeff_charm} shows that $c_{002}$
becomes slightly negative for temperatures up to about 220 MeV.  This
behavior is obviously unphysical, since this coefficient is directly
proportional to the necessarily positive charm quark number
susceptibility at zero chemical potential
$\chi_{cc}(\mu_{l,s,c}=0)\sim\VEV{n_c^2}$, where $n_c$ is the charm
quark number density. It follows that $c_{002}$ should be a
non-negative number at all temperatures.  We tracked this unphysical
behavior to the interplay between the heavy-quark mass and the Naik
term in the asqtad action. In the tuning of the latter,
corrections proportional to $m_c^2$ were not included. 
It is easiest to understand this if we examine the 
quark number susceptibility for free asqtad (Naik) fermions at large quark masses
shown in Fig.~\ref{fig:free_Iisentr_charm} (left panel). 
In the continuum limit, this susceptibility should approach zero from above with increasing
heavy-quark mass. We find that at $N_t=6$ and $8$ there is
a pronounced ``dip'' into negative values for a certain range of large quark masses.
This effect is much smaller at $N_t=12$. Since this particular discretization effect
does not occur for standard staggered fermions at $N_t=6$, we conclude that 
certain thermodynamic quantities,
such as susceptibilities, are sensitive to the "length"  of the Naik term 
and require large $N_t$'s in order to overcome their unphysical behavior.  
From Fig.~\ref{fig:free_Iisentr_charm} (left panel), the p4 action seems to be
much closer to the continuum limit at $N_t=6$ and very probably will not show
this particular discretization effect in the dynamical case. 
The HISQ action \cite{Follana:2006rc} improves the heavy-quark dispersion
relation by tuning the coefficient of the Naik term.  (The same tuning could
have been done with the asqtad action.)  Tuning suppresses this
unphysical behavior for $N_t \ge 6$ for the range of $m/T$ up to $O(8)$. 
Still, in our 
unquenched 2+1 flavor case, the negative dip in the
$c_{002}$ coefficient is quite small, so that its effect, for example, on the isentropic EOS 
is negligible
over the parameter range relevant to heavy-ion collisions.
 Of course, other mixed and unmixed coefficients might be affected by the
limited temporal extent $N_t=6$ as well, but since they are even smaller than $c_{002}$ we 
can also ignore their unphysical  contribution at low temperatures and small chemical potentials.
\begin{figure}[t]
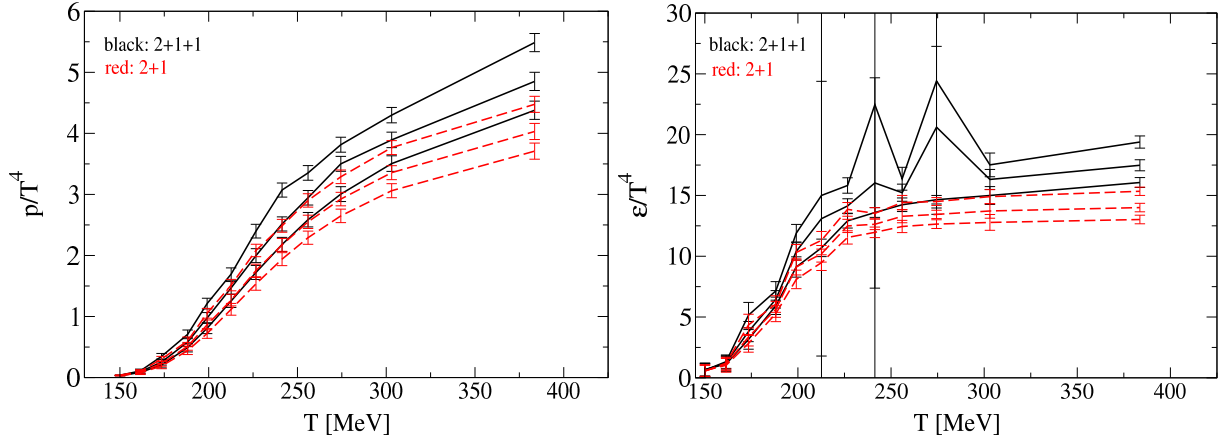

 \includegraphics*[width=8cm]{P_isentr.eps}
 \includegraphics*[width=8cm]{E_isentr.eps}
\caption{(Left panel) The isentropic pressure at selected $s/n_B$ values 
for 2+1 and 2+1+1 flavors (red and black respectively). The data ordering
is as in the right pannel of Fig.~\ref{fig:free_Iisentr_charm}.
(Right panel)
The same for the isentropic energy density. 
}
\label{fig:pe_isentr_charm}
\end{figure}

From the point of view of the isentropic EOS, our results show that the effect of the charm quark
cannot be simply ignored. We have determined the approximate isentropic trajectories in 
the ($\mu_l$, $\mu_s$, $\mu_c$, $T$)
space, 
by numerically solving the system
\be
{s\over n_B}(\mu_l, \mu_s, \mu_c,T)= C,\hspace{0.5cm}
{n_s\over T^3} (\mu_l, \mu_s, \mu_c,T) = 0,\hspace{0.5cm}
{n_c\over T^3}(\mu_l, \mu_s, \mu_c,T) = 0,
\ee
with $C=30$, 45, and 300. Figures~\ref{fig:free_Iisentr_charm} (right panel)~and~\ref{fig:pe_isentr_charm} (both panels)
 present the 2+1+1 flavor isentropic interaction measure,
pressure, and energy density, respectively, and compare them with the 2+1 flavor case.
We see that the charm quark contribution is non-negligible,
although it is due
mainly to the contribution of the zeroth order coefficients in the 
Taylor expansions ({\it i.e.}, the EOS calculated at zero chemical potential). We also note that
for the range of temperatures between about 220 and 280 MeV, the errors on the isentropic 
interaction measure become large. In this region of the isentropic 
trajectory, $\mu_c$ is big enough to make contributions from the quite noisy mixed
coefficients visible. The isentropic energy density, of course, inherits this feature, being 
a linear combination of the interaction measure and the pressure.

\section{Effects of the bottom quark on the EOS}

\begin{figure}[t]
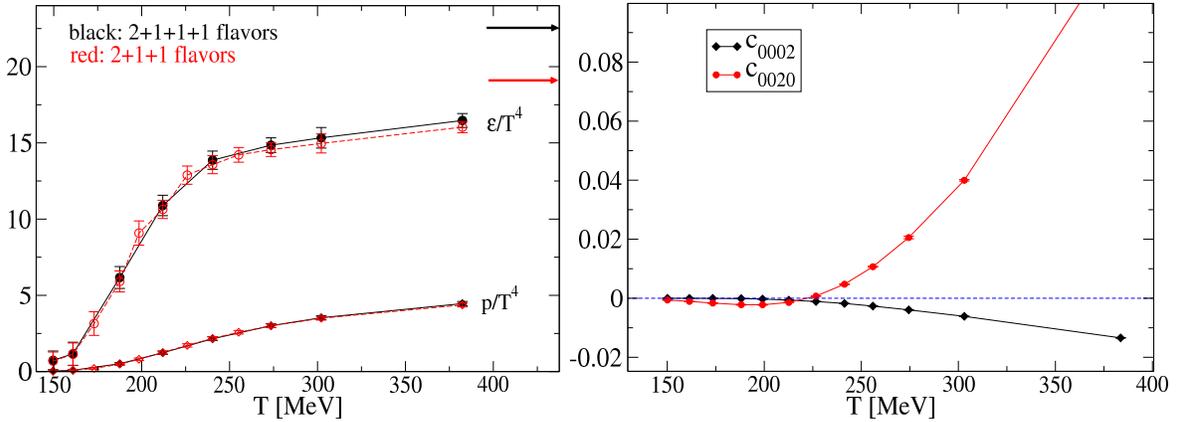

  \includegraphics*[width=7.4cm]{EOS_charm_bot.eps}
  \includegraphics*[width=8cm]{charm_bot_coeff.eps}
\caption{(Left panel) The pressure and energy density {\it vs.} temperature
for 2+1+1 and 2+1+1+1 flavors (red and black, respectively). The arrows indicate the
energy density Stefan-Boltzmann values for both cases.
(Right panel) The pressure Taylor expansion coefficients $c_{0020}$ and $c_{0002}$
{\it vs.} temperature.
}
\label{fig:EOS_cb}
\end{figure}

In the previous section, we presented evidence that the charm quark contributions 
to the EOS are non-negligible. At still higher temperatures the $b$- and eventually $t$-quark
contributions should be similarly non-negligible.
In this section, we examine the effects of the bottom quark on the EOS
in the range of temperatures up to about 400 MeV. Since the bottom quark is considerably 
heavier than the charm quark, we expect its contribution to the EOS to be smaller.
To estimate it, we simply repeated the charm quark calculation but with a heavier
mass corresponding to the bottom quark. The quenching error, even if relevant, will be 
smaller than the corresponding one for the charm quark.
On the ensemble that we used for the charm quark, we tune the
bottom quark mass to match the $B_s$ rest mass to its experimental
value. We found that within 5\% $m_b/m_s=38$.  We kept
that ratio constant along the physics trajectory. The problems of the
tuning of the bottom quark are potentially worse than in the case of
the charm quark, but we do not expect them to skew significantly our
final result for the EOS, since the bottom quark contribution itself is expected to
be small.
Figure~\ref{fig:EOS_cb} (left panel) shows the pressure and energy density at zero chemical potential
with ($2+1+1+1$) and without ($2+1+1$) the bottom quark.
We can conclude that the bottom quark contribution to the EOS at zero chemical 
potential is small (less than a standard deviation) 
in the transition region. It grows to about a standard deviation at temperatures close to 400 MeV.
However, the range of temperatures we examine here is somewhat limited and probably by $T\sim600$ MeV
the bottom quark effects would grow to be statistically significant for comparable 
statistics at that temperature. We also have to bear in mind that the heavy-quark discretization effects
may play a significant role here and keep the bottom quark contribution lower than what
it would be in the continuum limit. 

As for the EOS at small nonzero chemical potential, our results for
the coefficients of the Taylor series beyond the zeroth order term
discussed above show that the bottom quark contribution can be safely
ignored at the present level of statistics.
The discretization effect which we
found for the charm quark in the previous section is much worse for the bottom quark. 
Figure~\ref{fig:EOS_cb} (right panel)
compares the $c_{0020}$ coefficient (referred to as $c_{002}$ in the previous section) and 
the coefficient $c_{0002}$ in the Taylor expansion for the pressure when all chemical potentials
$\mu_{l,s,c,b}\neq 0$. The $c_{0002}$ is persistently negative at all available temperatures.
At the present level of statistics and small nonzero chemical potential, this
effect is unimportant.  But if higher precision is desired, one should
tune the coefficient of the Naik term and increase $N_t$.

\section{Conclusions}
We extended our thermodynamics study of the quark-gluon plasma 
with chemical potential to 
finer lattices with temporal extent 
$N_t=6$.
Comparing our results with previous results at $N_t = 4$ gives an indication
of the importance of cutoff effects.  As before,
we used the Taylor expansion method to sixth order
for the case of 2+1 quark flavors.
We found small but significant changes in the coefficients of the
Taylor expansions of the pressure and interaction measure in going
from $N_t=4$ to $6$.
This leads to small differences in the resulting interaction
measure, pressure, and energy density between the two cases,
when matching to the experimental condition of zero strange quark density
and keeping $\bar\mu_l/T$ constant. 
Under these conditions, 
small discretization effects are also
visible in the light-light, light-strange and strange-strange quark number susceptibilities
and the light-quark density.
On the other hand, the isentropic EOS shows very little difference between $N_t=4$ and $6$.
More pronounced lattice spacing effects are evident in the isentropic light-light
and strange-strange quark number susceptibilities, which we attribute to the fact that 
these quantities have contributions only from the nonzeroth order 
Taylor expansion coefficients which are more sensitive to the cutoff. 
And finally, we did not find any peaks 
along the isentropic trajectories, which suggests that current experiments
operate away from a possible critical point.

A full-flavor quark-gluon plasma EOS is undoubtedly important for cosmological studies.
Accordingly, we determined 
the effects of the charm quark (at zero and nonzero chemical potential) and 
the bottom quark (at zero chemical potential only) on the EOS. 
Both heavy quarks were represented in the 
heavy-quark-quenched 
approximation by asqtad valence quarks.
We expect that the quenching error for such heavy quarks is small, especially for the $b$ quark,
but only a direct
comparison with a calculation with dynamical $c$ and $b$ quarks can confirm that. 
We found that the contribution of the charm quark at zero chemical potential reaches about 20\%
in the energy density at temperatures of about 400 MeV and  cannot be ignored 
in a high-precision cosmological calculation of the properties of the early Universe. The bottom
quark contribution is within a standard deviation at that temperature. Our results for 
the charm and bottom effects on the EOS, however,
may be affected by the heavy-quark discretization error we find in the free asqtad action
calculation. This implies that they are possibly lower than their respective continuum values.

At nonzero chemical potential, both charm and bottom quarks present a problem (the bottom
quark much more so),
since we found heavy-quark discretization effects in the Taylor expansion coefficients
(especially large for the bottom quark),
which could be overcome by tuning the coefficient of the Naik term
and/or using  $N_t>6$. However, the charm and bottom quark 
contributions to the EOS due exclusively to the nonzero chemical potential are very small
over the parameter range accessible to heavy-ion collisons,
and at our level of precision they are entirely within the present
statistical errors of the EOS at zero chemical potential.

\section*{
ACKNOWLEDGMENTS}
This work was supported by the U.S. Department of Energy under grant
numbers
DE-FC02-06ER-41439, 
DE-FC02-06ER-41443, 
DE-FC06-01ER-41437, 
DE-FG02-04ER-41298, 
and
DE-FG02-91ER-40661  
and by the U.S. National Science Foundation under grant numbers
OCI08-32315, 
PHY05-55234, 
PHY05-55397, 
PHY07-03296, 
PHY07-57035, 
PHY07-57333, 
PHY07-04171, 
and          %
PHY09-03536. 
An allocation of computer time from the Center for High Performance
Computing at the University of Utah is gratefully acknowledged. Code
development was carried out in part using the computational resources
of Indiana University.  Computation for this research was supported in
part by the U.S. National Science Foundation through TeraGrid
resources provided by the Texas Advanced Computing Center (TACC), the
Louisiana Optical Network Initiative (LONI), and the National Center
for Supercomputing Applications (NCSA) under grant number
TG-MCA93S002. Computation for this work was also carried out on the
Fermilab LQCD cluster, supported by the Offices of Science, High
Energy Physics, and Nuclear Physics of the U.S. Department of Energy.



\appendix
\section{General framework for adding the charm quark to the EOS at nonzero
chemical potential}
With the addition of the charm quark to the $u$, $d$ and $s$
quarks in the sea, the partition function becomes: 
\be
{\cal Z} = \int {\cal D}U\, e^{\frac{n_l}{4}\ln {\rm det}\,{\cal M}_l}
 e^{\frac{n_s}{4} \ln {\rm det}\,{\cal M}_s}
 e^{\frac{n_c}{4} \ln {\rm det}\,{\cal M}_c} e^{-S_g},
\ee
with ${\cal M}_f$ being the quark matrix for flavor $f$.
Thus, the pressure can be now expanded in the following manner:
\begin{equation}
{p\over T^4}=
\sum_{n,m,k=0}^\infty c_{nmk}(T) \left({\bar{\mu}_l\over T}\right)^n
\left({\bar{\mu}_s\over T}\right)^m\left({\bar{\mu}_c\over T}\right)^k,
\label{eq:p}
\end{equation}
where $\bar{\mu}_f$ is the quark chemical potential  for 
flavor $f$ and the coefficients are
\begin{equation}
c_{nmk} (T)=
{1\over n!}{1\over m!}{1\over k!}{N_\tau^{3}\over N_\sigma^3}{{\partial^{n+m+k}
\ln{\cal Z}}\over{\partial(\mu_l N_\tau)^n}{\partial(\mu_s N_\tau)^m}
{\partial(\mu_c N_\tau)^k}}\biggr\vert_{\mu_{l,s,c}=0},
\label{eq:cn}
\end{equation}
with $\mu_f$ is the quark chemical potential in lattice units.
The coefficients above are nonzero only if $n+m+k$ is even.
A similar expansion applies to the interaction measure. These
coefficients are for the asqtad quark action:
\bea
\hspace{-2cm}b_{nmk} &=& -{1\over n!m!k!}{N_t^3\over N_s^3}
\sum_{f=l,s,c}\frac{n_f}{4}
\left[
\left.\frac{ d(m_fa)}{d\ln a}\right|_{\mu_{l,s,c}=0}{\rm tr}\left.{\partial^{n+m+k} 
\langle 2M_f^{-1}\rangle\over \partial(\mu_l N_t)^n
\partial(\mu_s N_t)^m\partial(\mu_c N_t)^k}\right|_{\mu_{l,s,c}=0}\right.\nn\\
&&+
\left.\left.\frac{du_0}{d\ln a}\right|_{\mu_{l,s,c}=0}{\rm tr}\left.{\partial^{n+m+k}
\langle M_f^{-1}\frac{dM_f}{du_0}\rangle
\over \partial(\mu_l N_t)^n\partial(\mu_s N_t)^m\partial(\mu_c N_t)^k}\right|_{\mu_{l,s,c}=0}
\right]\nn\\
&& -{1\over n!m!k!}{N_t^3\over N_s^3}\left.{\partial^{n+m+k}\langle {\cal G}\rangle
\over \partial(\mu_l N_t)^n\partial(\mu_s N_t)^m\partial(\mu_c N_t)^k}\right|_{\mu_{l,s,c}=0}.
\eea
In the above, ${\cal G}=-dS_g/d\ln a$, with $S_g$ being the gluon part of the action. 
\section{Calculating the pressure coefficients in the Taylor expansion}
These are most easily calculated using the following (similar 
to the derivation in the Appendix in Ref.~\cite{Bernard:2007nm}):
\bea
\frac{\partial\,\ln{\cal Z}}{\partial \mu_l}&\equiv&\A_{100} = \langle L_1\rangle,\\
\frac{\partial\,\ln{\cal Z}}{\partial \mu_s}&\equiv&\A_{010} = \langle H_1\rangle,\\
\frac{\partial\,\ln{\cal Z}}{\partial \mu_c}&\equiv&\A_{001} = \langle Q_1\rangle.
\eea
It can be shown that
\bea
\frac{\partial\A_{nmk}}{\partial \mu_l}&=& \A_{n+1,m,k} - \A_{100}\A_{nmk},\\
\frac{\partial\A_{nmk}}{\partial \mu_s}&=& \A_{n,m+1,k} - \A_{010}\A_{nmk},\\
\frac{\partial\A_{nmk}}{\partial \mu_c}&=& \A_{n,m,k+1} - \A_{001}\A_{nmk},
\eea
where
\be
\A_{nmk}\equiv\left\langle e^{-L_{0}} e^{-H_{0}} e^{-Q_{0}}
\frac{\partial^n  e^{L_{0}}}{\partial\mu_l^n}
\frac{\partial^m  e^{H_{0}}}{\partial\mu_s^m}
\frac{\partial^k e^{Q_{0}}}{\partial\mu_c^k}\right\rangle.
\label{eq:A}
\ee
In all of the above, $Q_k$ is defined as:
\be
Q_k = \frac{n_c}{4} \frac{\partial^k \ln \det M_c}{\partial \mu_c^k}.
\ee
The operators $L_k$ and $H_k$ have a similar form for the light and the strange quark respectively.
All coefficients $c_{nmk}$ which have at least one of the indices
equal to zero have the same form as in the Appendix of Ref.~\cite{Bernard:2007nm}, 
with appropriate substitutions of $L_n$ or $H_n$ with $Q_k$.
The ``new'' coefficients that appear to $O(6)$ are 
$c_{(2,1,1)}$, $c_{(3,2,1)}$, $c_{222}$ and $c_{(4,1,1)}$,
where the notation $(m,n,k)$ means all distinct permutations
of the indices. Explicitly we have
\bea
c_{211} &=& \frac{1}{2!1!1!}\frac{1}{N_s^3N_t}(\A_{211}-2\A_{110}\A_{101} -\A_{011}\A_{200}),\\
c_{222} &=&\frac{1}{2!2!2!}\frac{1}{N_s^3N_t^3} 
(\A_{222}+16\A_{110}\A_{101}\A_{011} +4\A_{101}^2\A_{020}
+4\A_{011}^2\A_{200}+4\A_{110}^2\A_{002} +2\A_{002}\A_{200}\A_{020}\nn\\
&&-\A_{002}\A_{220}-\A_{200}\A_{022}-\A_{202}\A_{020}
-4\A_{101}\A_{121}-4\A_{110}\A_{112}-4\A_{011}\A_{211}), \\
c_{321} &=&\frac{1}{3!2!1!}\frac{1}{N_s^3N_t^3}
(\A_{321}+12\A_{110}^2\A_{101}-6\A_{211}\A_{110}
-2\A_{011}\A_{310}-3\A_{220}\A_{101}-\A_{020}\A_{301}\nn\\
&&-3\A_{200}\A_{121}+6\A_{020}\A_{200}\A_{101}+12\A_{011}\A_{110}\A_{200}),\\
c_{411} &=& \frac{1}{4!1!1!}\frac{1}{N_s^3N_t^3}
(\A_{411}+ 24\A_{110}\A_{101}\A_{200}-4\A_{301}\A_{110}-4\A_{101}\A_{310}
-\A_{011}\A_{400}\nn\\
&&-6\A_{200}\A_{211}+6\A_{011}\A_{200}^2).
\eea
Permuting the indices above gives us the rest of the coefficients.
Calculating the $\A_{nmk}$ is straightforward from Eq.~(\ref{eq:A}).

\section{Calculating the interaction measure coefficients in the Taylor expansion}
\subsection{First type of derivative}
This section gives a method to calculate the derivative:
\be
 \left.{\partial^{n+m+k} \langle M_f^{-1}\rangle\over \partial(\mu_l N_t)^n
\partial(\mu_s N_t)^m\partial(\mu_c N_t)^k}\right|_{\mu_{l,s,c}=0},
\ee
for $f=l,s,c$, when all of the indices $n,m$ and $k$ are nonzero. See Ref.~\cite{Bernard:2007nm} for 
results when at least one is zero.
It is convenient to define the observables:
\bea
\label{eq:b}
\B_{nmk} &\equiv&\left\langle {\rm e}^{-L_{0}}{\rm e}^{-H_{0}}{\rm e}^{-Q_{0}}
\frac{\partial^n ({\rm tr}\, M_l^{-1}{\rm e}^{L_{0}})}{\partial\mu_l^n}
\frac{\partial^m {\rm e}^{H_{0}}}{\partial\mu_s^m}
\frac{\partial^k {\rm e}^{Q_{0}}}{\partial\mu_c^k}\right\rangle,\\
\B^\prime_{nmk} &\equiv&\left\langle {\rm e}^{-L_{0}}{\rm e}^{-H_{0}}{\rm e}^{-Q_{0}}
\frac{\partial^n {\rm e}^{L_{0}}}{\partial\mu_l^n}
\frac{\partial^m ({\rm tr}\, M_s^{-1}{\rm e}^{H_{0}})}{\partial\mu_s^m}
\frac{\partial^k {\rm e}^{Q_{0}}}{\partial\mu_c^k}\right\rangle,\\
\B^{\prime\prime}_{nmk} &\equiv&\left\langle {\rm e}^{-L_{0}}{\rm e}^{-H_{0}}{\rm e}^{-Q_{0}}
\frac{\partial^n {\rm e}^{L_{0}}}{\partial\mu_l^n}
\frac{\partial^m {\rm e}^{H_{0}}}{\partial\mu_s^m}
\frac{\partial^k ({\rm tr}\, M_c^{-1}{\rm e}^{Q_{0}})}{\partial\mu_c^k}\right\rangle.
\eea
The above means:
\bea
\B_{000}&\equiv&\left\langle{\rm tr}\, M_l^{-1}\right\rangle,\\
\B^\prime_{000}&\equiv&\left\langle{\rm tr}\, M_s^{-1}\right\rangle,\\
\B^{\prime\prime}_{000}&\equiv&\left\langle{\rm tr}\, M_c^{-1}\right\rangle.
\eea
Let $f=l$ then we have the following rule:
\bea
\frac{\partial \B_{nmk}}{\partial \mu_l}& =& \B_{n+1,mk} - \A_{100}\B_{nmk},\\
\frac{\partial \B_{nmk}}{\partial \mu_s}& =& \B_{n,m+1,k} - \A_{010}\B_{nmk},\\
\frac{\partial \B_{nmk}}{\partial \mu_c}& =& \B_{nm,k+1} - \A_{001}\B_{nmk}.
\eea
Using the above, we calculate:
\bea
\frac{\partial^4\left\langle {\rm tr}M_l^{-1} \right\rangle}{\partial^2 \mu_l\partial \mu_s\partial \mu_c } &=& 
\B_{211}+ 2\B_{000}\A_{011}\A_{200}-\B_{000}\A_{211}
+4\B_{000}\A_{110}\A_{101}-\A_{011}\B_{200}\nn\\
&&-2\B_{101}\A_{110}-2\B_{110}\A_{101}-\B_{011}\A_{200},\\
\frac{\partial^6\left\langle {\rm tr}M_l^{-1} \right\rangle}{\partial^2 \mu_l\partial^2 \mu_s\partial^2 \mu_c } &=& 
 \B_{222} -\B_{000}\A_{222} +2\A_{002}\A_{200}\B_{020} +2\A_{002}\B_{200}\A_{020}+2\B_{002}\A_{200}\A_{020}\nn\\
&&-4\A_{011}\B_{211} -4\B_{112}\A_{110} -4\A_{101}\B_{121} -\B_{002}\A_{220}-\B_{200}\A_{022}-\A_{202}\B_{020}\nn\\
&&+16\A_{011}\A_{101}\B_{110} +16\A_{011}\B_{101}\A_{110} +16\B_{011}\A_{101}\A_{110}\nn\\
&&+8\B_{000}\A_{211}\A_{011}+8\B_{000}\A_{101}\A_{121}+8\B_{000}\A_{110}\A_{112} \nn\\
&&+8\B_{101}\A_{101}\A_{020} +8\A_{002}\A_{110}\B_{110}+8\A_{011}\A_{200}\B_{011}\nn\\
&& +2\B_{000}\A_{202}\A_{020}+2\A_{002}\B_{000}\A_{220}+2\B_{000}\A_{200}\A_{022}\nn\\
&& -12\A_{002}\B_{000}\A_{110}^2-12\B_{000}\A_{200}\A_{011}^2-12\B_{000}\A_{101}^2\A_{020}\nn\\
&&-\A_{002}\B_{220}-\A_{200}\B_{022}-\B_{202}\A_{020}-4\B_{101}\A_{121}-4\B_{110}\A_{112}-4\B_{011}\A_{211}\nn\\
&&+4\B_{002}\A_{110}^2+4\A_{011}^2\B_{200}+4\A_{101}^2\B_{020}\nn\\
&& -6\A_{002}\B_{000}\A_{200}\A_{020}-48\B_{000}\A_{110}\A_{011}\A_{101},\\
\frac{\partial^6\left\langle {\rm tr}M_l^{-1} \right\rangle}{\partial^3 \mu_l\partial^2 \mu_s\partial^1 \mu_c } &=& 
\B_{321}-\B_{000}\A_{321}+ 12\A_{110}\A_{011}\B_{200}+6\B_{000}\A_{220}\A_{101}+12\B_{110}\A_{011}\A_{200}\nn\\
&&-36\B_{000}\A_{110}^2\A_{101}+12\B_{011}\A_{200}\A_{110}+6\A_{020}\A_{200}\B_{101}+24\B_{110}\A_{101}\A_{110}\nn\\
&&-6\A_{110}\B_{211}-3\B_{200}\A_{121}+2\B_{000}\A_{020}\A_{301}+12\A_{110}^2\B_{101}+12\B_{000}\A_{110}\A_{211}\nn\\
&&-3\A_{220}\B_{101}+6\B_{000}\A_{200}\A_{121}-6\A_{211}\B_{110}-3\B_{220}\A_{101}-2\B_{011}\A_{310}\nn\\
&&-\A_{020}\B_{301}-\B_{020}\A_{301}-36\A_{011}\B_{000}\A_{110}\A_{200}-3\A_{200}\B_{121}+6\A_{020}\A_{101}\B_{200}\nn\\
&&-18\A_{020}\B_{000}\A_{200}\A_{101}+4\A_{011}\B_{000}\A_{310}+6\A_{101}\B_{020}\A_{200}-2\A_{011}\B_{310},\\
\frac{\partial^6\left\langle {\rm tr}M_l^{-1} \right\rangle}{\partial^4 \mu_l\partial^1 \mu_s\partial^1 \mu_c } &=& 
\B_{411}-\B_{000}\A_{411} -\B_{011}\A_{400}-\A_{011}\B_{400}+ 24\A_{101}\B_{110}\A_{200}+24\A_{110}\A_{200}\B_{101}\nn\\
&&+12\A_{200}\A_{011}\B_{200}+24\A_{110}\A_{101}\B_{200}+12\B_{000}\A_{200}\A_{211}\nn\\
&&+8\B_{000}\A_{301}\A_{110}+8\A_{101}\B_{000}\A_{310}-4\A_{101}\B_{310}-4\A_{110}\B_{301}\nn\\
&&-4\B_{110}\A_{301}-4\A_{310}\B_{101}-6\A_{200}\B_{211}-6\A_{211}\B_{200}-72\B_{000}\A_{200}\A_{101}\A_{110}\nn\\
&&+6\A_{200}^2\B_{011}+2\B_{000}\A_{011}\A_{400}-18\A_{011}\A_{200}^2\B_{000}.
\eea
Replacing $\B$ with $\B^\prime$ or $\B^{\prime\prime}$ in the above, we get
the expressions for the derivatives of
$\left\langle {\rm tr}M_s^{-1} \right\rangle$ or $\left\langle {\rm tr}M_c^{-1} \right\rangle$.
The explicit forms of $\B_{nmk}$ are easy to deduce from Eq.~(\ref{eq:b}).
To get the $\B^\prime_{nmk}$ or $\B^{\prime\prime}_{nmk}$ we need to interchange appropriately
the three observables:

\bea
{\el}_n &=& \frac{\partial^n {\rm tr}\,M_l^{-1}}{\partial \mu_l^n},\\
\h_n &=& \frac{\partial^n {\rm tr}\,M_s^{-1}}{\partial \mu_s^n},\\
q_n &=& \frac{\partial^n {\rm tr}\,M_c^{-1}}{\partial \mu_c^n},
\eea
along with $L_n$, $H_n$ and $Q_n$ in the explicit forms of $\B_{nmk}$.

\subsection{Second type of derivative}
We also need to calculate the derivatives:
\be
\left.{\partial^{n+m+k}
\langle M_f^{-1}\frac{dM_f}{du_0}\rangle
\over \partial(\mu_l N_t)^n\partial(\mu_s N_t)^m\partial(\mu_c N_t)^k}\right|_{\mu_{l,s,c}=0},
\ee
where again $f=l,s,c$.
Similarly to the previous subsection, we define the observables:
\bea
\C_{nmk} &\equiv&\left\langle {\rm e}^{-L_{0}}{\rm e}^{-H_{0}}{\rm e}^{-Q_{0}}
\frac{\partial^n [{\rm tr}\, (M_l^{-1}\frac{dM_l}{du_0}){\rm e}^{L_{0}}]}{\partial\mu_l^n}
\frac{\partial^m {\rm e}^{H_{0}}}{\partial\mu_s^m}
\frac{\partial^k {\rm e}^{Q_{0}}}{\partial\mu_c^k}\right\rangle,\\
\C^\prime_{nmk} &\equiv&\left\langle {\rm e}^{-L_{0}}{\rm e}^{-H_{0}}{\rm e}^{-Q_{0}}
\frac{\partial^n {\rm e}^{L_{0}}}{\partial\mu_l^n}
\frac{\partial^m [{\rm tr}\, (M_s^{-1}\frac{dM_s}{du_0}){\rm e}^{H_{0}}]}{\partial\mu_s^m}
\frac{\partial^k {\rm e}^{Q_{0}}}{\partial\mu_c^k}\right\rangle,\\
\C^{\prime\prime}_{nmk} &\equiv&\left\langle {\rm e}^{-L_{0}}{\rm e}^{-H_{0}}{\rm e}^{-Q_{0}}
\frac{\partial^n {\rm e}^{L_{0}}}{\partial\mu_l^n}
\frac{\partial^m {\rm e}^{H_{0}}}{\partial\mu_s^m}
\frac{\partial^k [{\rm tr}\, (M_c^{-1}\frac{dM_c}{du_0}){\rm e}^{Q_{0}}]}{\partial\mu_c^k}\right\rangle.
\eea
From the above,
\bea
\C_{000}&\equiv&\left\langle{\rm tr}\, (M_l^{-1}\frac{dM_l}{du_0})\right\rangle,\\
\C^{\prime}_{000}&\equiv&\left\langle{\rm tr}\,( M_s^{-1}\frac{dM_s}{du_0})\right\rangle,\\
\C^{\prime\prime}_{000}&\equiv&\left\langle{\rm tr}\,( M_c^{-1}\frac{dM_c}{du_0})\right\rangle \, .
\eea
Let $f=l$, then it is easy to see that
\bea
\frac{\partial \C_{nmk}}{\partial \mu_l}& =& \C_{n+1,mk} - \A_{100}\C_{nmk},\\
\frac{\partial \C_{nmk}}{\partial \mu_s}& =& \C_{n,m+1,k} - \A_{010}\C_{nmk},\\
\frac{\partial \C_{nmk}}{\partial \mu_c}& =& \C_{nm,k+1} - \A_{001}\C_{nmk}.
\eea
Similar expressions apply in the case of $\C_{nmk}^\prime$ and $\C_{nmk}^{\prime\prime}$.
Then the derivatives
\be
\frac{\partial^n \left\langle{\rm tr}\, (M_{l,s,c}^{-1}\frac{dM_{l,s}}{du_0})\right\rangle}{\partial\mu_{l,s,c}^n}
\ee
have the form of the derivatives of $\left\langle{\rm tr}\, (M_{l,s,c}^{-1})\right\rangle$ 
in the previous section with the substitutions
$\B_{nmk}\rightarrow \C_{nmk}$, $\B_{nmk}^\prime\rightarrow \C_{nmk}^\prime$ and 
$\B_{nmk}^{\prime\prime}\rightarrow \C_{nmk}^{\prime\prime}$.
The explicit forms of $\C_{nmk}$, $\C_{nmk}^\prime$ and $\C_{nmk}^{\prime\prime}$
are the same as for $\B_{nmk}$, $\B_{nmk}^{\prime}$ and $\B_{nmk}^{\prime\prime}$ with the substitutions
$l_n\rightarrow\lambda_{n}$, $h_n\rightarrow\chi_n$ and $q_n\rightarrow\eta_n$ where
\bea
\lambda_n &=& \frac{\partial^n {\rm tr}\, (M_{l}^{-1}\frac{dM_{l}}{du_0})}{\partial\mu_{l}^n} \, ,\\
\chi_n &=& \frac{\partial^n {\rm tr}\, (M_{s}^{-1}\frac{dM_{s}}{du_0})}{\partial\mu_{s}^n},\\
\eta_n &=& \frac{\partial^n {\rm tr}\, (M_{c}^{-1}\frac{dM_{c}}{du_0})}{\partial\mu_{c}^n}.
\eea

\subsection{Third type of derivative}
The last type of derivative that we need is the gauge derivative
\be
\left.{\partial^{n+m+k}\langle {\cal G}\rangle
\over \partial(\mu_l N_t)^n\partial(\mu_s N_t)^m\partial(\mu_c N_t)^k}\right|_{\mu_{l,s,c}=0}.
\ee
In this case, let
\be
G_{nmk}\equiv\left\langle {\cal G}\,{\rm e}^{-L_{0}}{\rm e}^{-H_{0}}{\rm e}^{-Q_{0}}
\frac{\partial^n {\rm e}^{L_{0}}}{\partial\mu_l^n}
\frac{\partial^m {\rm e}^{H_{0}}}{\partial\mu_s^m}
\frac{\partial^p {\rm e}^{Q_{0}}}{\partial\mu_c^k}\right\rangle,
\ee
and similarly as before
\bea
\frac{\partial G_{nmk}}{\partial \mu_l}&=& G_{n+1,mk} - \A_{100}G_{nmk},\\
\frac{\partial G_{nmk}}{\partial \mu_s}&=& G_{n,m+1,k} - \A_{010}G_{nmk},\\
\frac{\partial G_{nmk}}{\partial \mu_c}&=& G_{nm,k+1} - \A_{001}G_{nmk},
\eea
with
\be
G_{000} = \langle {\cal G}\rangle.
\ee
This means that the necessary derivatives $\left.{\partial^{n+m+k}\langle {\cal G}\rangle
\over \partial(\mu_l N_t)^n\partial(\mu_s N_t)^m\partial(\mu_c N_t)^k}\right|_{\mu_{l,s,c}=0}$ 
have the same form as the derivatives $\left.{\partial^{n+m+k}{\rm tr}\langle M_f^{-1}\rangle\over 
\partial(\mu_l N_t)^n\partial(\mu_s N_t)^m\partial(\mu_c N_t)^k}\right|_{\mu_{l,s,c}=0}$ 
with $\B_{nmk}\rightarrow G_{nmk}$.
The $G_{nmk}$ observables have very similar form to the $\A_{nmk}$ observables,
but with an additional multiplication by ${\cal G}$ inside the ensemble average brackets of
each term in them.

\end{document}